# PRACTICAL REALIZATION OF THE QUANTUM CRYPTOGRAPHY
# PROTOCOL EXPLOITING POLARIZATION ENCODING IN THE QUTRITS

**G.A.Maslennikov, A.A.Zhukov, M.V.Chekhova, S.P.Kulik**

*Chair of Quantum Electronics, Department of Physics,*
*Moscow M.V.Lomonosov State University,*
*Russia, Moscow 119992,*
*tel: 7 (095) 939 4372, fax: 7 (095) 939 1104, e-mail: postmast@qopt.phys.msu.su*

**Abstract**

We propose and discuss a specific scheme allowing to realize a Quantum Cryptography qutrit protocol. This protocol exploits the polarization properties of single-frequency and single-spatial mode biphotons.

**Introduction**

The art of quantum cryptography (QC) could well be the first practical application of Quantum Information at the single quanta level. It provides two distant users Alice and Bob with random secret key, which later can be used for encrypting messages in "one-time pad" scheme – the only cryptographic scheme that is mathematically proven to be secure. The simplest method is the use of polarized single photons as the quantum carriers between Alice and Bob. The orthogonal polarization states represent bit values 0 and 1. The first QC protocol, based on such systems was described by Bennett and Brassard [1]. It requires the use of four qubits prepared in four different states that belong to two mutually unbiased bases. It means that any state vectors $|e_i\rangle, |e_j\rangle$ that belong to different bases must satisfy the following condition $\left|\langle e_i | e_j \rangle\right|^2 = \frac{1}{2}$. One may also extend this protocol, which is commonly called BB84, to the case where three mutually unbiased bases are used. This extension leads to the improvement of the security against eavesdropping in the case of so-called symmetric attacks [2, 3]. Various methods are used to implement these protocols in experiment (for details, see the review [4]) and a lot of experiments were carried out in the last years. It is important to mention that some of the experiments also show excellent results outside the lab in the real surroundings [5]. Possible kinds of eavesdropper's attacks on these protocols are reviewed in [2-4, 6]. But even with present technologies, the most promising QC systems suffer from a low bit rate of some hundred Hz, compared to thousands of MHz achieved in classical systems. There are some ways to enhance the bit rate like the use of the single photon sources ("photon guns"), when one can be sure that there is only one photon emitted at a time, improvement of the detector's efficiency or by using higher-dimensional systems than qubits. In the last case, the amount of information that is carried will be proportional to $d^n$, where $d$ is the dimension of the system and $n$ the total number of systems. The possibility of increasing the system's dimension was proposed in [7]. The authors used four-level systems (qu-quarts) as an example and it was shown that the key generation rate is higher then in the case of qubits, and their protocol is also more sensitive to intercept-resend eavesdropping strategy. The other example of extending the system's space is the use of the three-level systems (qutrits); a QC protocol that exploits such systems was introduced in [8]. This protocol uses 12 states and four mutually unbiased bases to encode the information. The optimal eavesdropping on this protocol was studied in [9]. The entanglement based protocol for qutrits was also suggested recently and found to be more robust against the quantum cloning machines attacks [10, 11].

In this paper we will discuss a proposal for possible optical realization of the three-level systems, based on the polarization state of the biphotons – superposition of two-photon Fock states and vacuum and propose an experimental setup for the protocol suggested in [8].

**Biphotons as qutrits**

Several ways are known how to experimentally realize multilevel quantum optical systems. In one of them [7], the interferometric procedure of state preparation is used, when attenuated laser pulses are sent into a multi-arm interferometer. The number of arms is equal to the system's dimensionality. Identification of the basis states is done either through the pulses delay (temporal basis) or by the presence of constructive interference in a certain arm of the interferometer that is put into the registration system (energy basis). Another example is the optical field that consists of pairs of correlated photons (biphotons) that belong to different polarization modes. The preparation of such fields and their unitary transformations are described in [12, 13] and the measurement procedure (tomography) of such quantum objects can be found in [14]. The pure polarization state of a single-mode biphoton field (by saying "single-mode" we mean that photons that form a biphoton have equal frequencies and propagate along the same direction) can be written as

$$|\Psi\rangle = c_1|2,0\rangle + c_2|1,1\rangle + c_3|0,2\rangle = c_1|\boldsymbol{a}\rangle + c_2|\boldsymbol{b}\rangle + c_3|\boldsymbol{g}\rangle, \qquad (1)$$

where the first and the second position in brackets indicate the polarization modes (say, horizontal H and vertical V), the total number of photons in this mode is N = 2, and $c_i = |c_i| \exp(i f_i)$ are the complex probability amplitudes to find the biphoton in the corresponding state. The states $|2,0\rangle$ and $|0,2\rangle$ correspond to type-I phase-matching where both photons in a biphoton have the same polarization and the state $|1,1\rangle$ is obtained via type-II phase-matching where these photons are polarized orthogonally. One can see that the state $|\Psi\rangle$ describes a three-level system. Equation (1) presents three possible states of a biphoton written in the so-called "HV – basis". For the realization of the protocol that was proposed in [8] one needs three more bases that can be prepared from the natural HV - bases according to the rule

$$|\boldsymbol{a}'\rangle = \frac{1}{\sqrt{3}}\left(|\boldsymbol{a}\rangle + |\boldsymbol{b}\rangle + |\boldsymbol{g}\rangle\right),$$

$$|\boldsymbol{b}'\rangle = \frac{1}{\sqrt{3}}\left(|\boldsymbol{a}\rangle + \exp\left(\frac{2\boldsymbol{p}i}{3}\right)|\boldsymbol{b}\rangle + \exp\left(-\frac{2\boldsymbol{p}i}{3}\right)|\boldsymbol{g}\rangle\right), \qquad (2)$$

$$|\boldsymbol{g}'\rangle = \frac{1}{\sqrt{3}}\left(|\boldsymbol{a}\rangle + \exp\left(-\frac{2\boldsymbol{p}i}{3}\right)|\boldsymbol{b}\rangle + \exp\left(\frac{2\boldsymbol{p}i}{3}\right)|\boldsymbol{g}\rangle\right).$$

These states form the second basis and the two other bases can be taken as

$$|\boldsymbol{a}''\rangle = \frac{1}{\sqrt{3}}\left(\exp\left(\frac{2\boldsymbol{p}i}{3}\right)|\boldsymbol{a}\rangle + |\boldsymbol{b}\rangle + |\boldsymbol{g}\rangle\right),$$

$$|\boldsymbol{b}''\rangle = \frac{1}{\sqrt{3}}\left(|\boldsymbol{a}\rangle + \exp\left(\frac{2\boldsymbol{p}i}{3}\right)|\boldsymbol{b}\rangle + |\boldsymbol{g}\rangle\right), \qquad (3)$$

$$|\boldsymbol{g}''\rangle = \frac{1}{\sqrt{3}}\left(|\boldsymbol{a}\rangle + |\boldsymbol{b}\rangle + \exp\left(\frac{2\boldsymbol{p}i}{3}\right)|\boldsymbol{g}\rangle\right),$$

and

$$|\boldsymbol{a}'''\rangle = \frac{1}{\sqrt{3}}\left(\exp\left(-\frac{2\boldsymbol{p}i}{3}\right)|\boldsymbol{a}\rangle + |\boldsymbol{b}\rangle + |\boldsymbol{g}\rangle\right),$$

$$|\boldsymbol{b}'''\rangle = \frac{1}{\sqrt{3}}\left(|\boldsymbol{a}\rangle + \exp\left(-\frac{2\boldsymbol{p}i}{3}\right)|\boldsymbol{b}\rangle + |\boldsymbol{g}\rangle\right), \qquad (4)$$

$$|\boldsymbol{g}'''\rangle = \frac{1}{\sqrt{3}}\left(|\boldsymbol{a}\rangle + |\boldsymbol{b}\rangle + \exp\left(-\frac{2\boldsymbol{p}i}{3}\right)|\boldsymbol{g}\rangle\right).$$

All these bases are mutually unbiased. In the case of 3-dimensional systems it means that $\left|\langle e_i | e_j \rangle\right|^2 = \frac{1}{3}$. As it was shown in [15], the polarization state of the biphoton can be imaged on the surface of the Poincare sphere. The main idea is to image the two-photon polarization state as two points on the surface of the sphere, where each point corresponds

to the polarization state of a single photon from the pair and its position is described in a well-known technique. Sometimes such a mapping helps to understand better the polarization properties of single-mode biphotons. It follows from the representation of the state vector in the form

$$|\Psi\rangle = \frac{a^\dagger(\vartheta,\varphi)a^\dagger(\vartheta',\varphi')|vac\rangle}{\left\| a^\dagger(\vartheta,\varphi)a^\dagger(\vartheta',\varphi')|vac\rangle \right\|}, \tag{5}$$

where $a^\dagger(\vartheta,\varphi)$ and $a^\dagger(\vartheta',\varphi')$ are the creation and annihilation operators of the photon in the corresponding polarization mode, $a^\dagger(\vartheta,\varphi) = \cos\frac{\vartheta}{2}a_H^\dagger + e^{i\varphi}\sin\frac{\vartheta}{2}a_V^\dagger$, $a_{H,V}^\dagger$ are photon creation operators in horizontal (H) and vertical (V) polarization modes, $\varphi, \varphi' \in [0, 2\pi]$ and $\vartheta, \vartheta' \in [0, \pi]$ - are the azimuthal and polar angles that define the position of a photon on the Poincare sphere. The primed and unprimed indices correspond to different photons from the pair. In this case,

$$\varphi, \varphi' = \frac{\varphi_3}{2} \pm \frac{1}{2}\arccos\left[ \frac{|c_2|^2}{2|c_1||c_3|} - \sqrt{1 + \frac{|c_2|^4}{4|c_1|^2|c_3|^2} - \frac{|c_2|^2}{|c_1||c_3|}\cos(2\varphi_2 - \varphi_3)} \right], \tag{6}$$

$$\vartheta, \vartheta' = \arccos\left[ \frac{|c_1|^2 - |c_3|^2 \pm 2\sqrt{\left[|c_2|^2 - |c_1||c_3|\cos(2\varphi - \phi_3)\right]^2 - |c_1|^2|c_3|^2}}{1 + |c_2|^2 - 2|c_1||c_3|\cos(2\varphi - \phi_3)} \right]. \tag{7}$$

The polarization degree of a biphoton can be shown to be

$$P = \frac{2\cos\frac{a}{2}}{1 + \cos^2\frac{a}{2}}, \tag{8}$$

where $a$ is an angle between the lines that connect the position of each photon on the surface with the center (see Fig. 1a). It is important to note that using linear transformations one cannot change the polarization state of the biphoton to a state with different polarization degree. As it will be shown later this greatly reduces the possibility of the practical implementation of the discussed QC protocol. In Fig.1b all 12 states that are used in the protocol are shown on the Poincare sphere, according to (6, 7).

According to the protocol, Alice randomly chooses one of these four bases in which she will prepare her qutrit, then sends the qutrit to Bob, who measures it in a randomly chosen basis. After the bases reconciliation Alice and Bob will have the sifted key, from which they can distill the secret key by means of error correction and privacy amplification techniques.

**Preparation and measurement of qutrits in 4 given bases.**

*1. Preparation and transformation of qutrits (Alice's station).*

For the quantum key distribution it is necessary that Alice can send Bob one randomly prepared qutrit at a time and to establish good time synchronization to be sure that Bob measures the correct qutrit. We propose an interferometric setup, similar to one that was used in [13]. It is a Mach-Zehnder interferometer, with three arms and with nonlinear crystals of the corresponding type in each arm. In this setup one can generate truly arbitrary polarization state of the biphoton field, including those that are described by formulas (1-4). The selection between the twelve states is done by the phase shifters, which imply a certain phase shift to the states, according to Table 1. The possible practical realization of the interferometer is shown in Fig.2. To improve the stability of the interferometer and prevent the losses introduced by combining biphotons using beam-splitters, we propose to use the two-arm version where type-II nonlinear crystal is not located directly in the interferometer. The scheme works as follows: the laser beam is split on a non-symmetric beam-splitter. The transmitted part (66%) goes through the long arm of the interferometer and is used for pumping two type-I nonlinear crystals, whose axes are oriented orthogonally to each other. All crystals are oriented to produce the biphotons in the frequency-degenerate, collinear regime. In this arm we generate the superposition of the form

$$|c_1||2,0\rangle + e^{i f_{31}}|c_3||0,2\rangle, \tag{9}$$

where $f_{31}$ is the relative phase between the states $|2,0\rangle$ and $|0,2\rangle$. The amplitudes of this states can be changed by rotating a half-wave plate $\left(\lambda/2\right)_1$, located before the crystals, and the relative phase can be set by using the phase shifters (PS). A cut-off filter eliminates the pump. A moving mirror that is driven by a piezoelectric driver is used to apply a certain phase shift $f_2' = f_{31} - f_2$ between the superposition (9) and the state $|1,1\rangle$, which is obtained by pumping the type-II crystal with the pump from the short arm of the interferometer through the dichroic mirror (DM) that reflects the pump from the short arm and transmits the biphotons from the long arm. The amplitude $|c_2|$ is varied by means of $\left(\lambda/2\right)_2$ plate. Thus, at the output of the interferometer we have the coherent superposition that is given by (1) and by varying four parameters ($\left(\lambda/2\right)_1, \left(\lambda/2\right)_2, f_{31}$ and $f_2'$), we can generate all 12 states that are needed for the QC protocol.

We have to mention here that the proposed interferometric scheme allows one to introduce the timing using short-pulsed pump. In this regime approximately one pair of photons, i.e. a single qutrit is emitted at given time, determined by the repetition rate and pulse duration of the pump laser. Possible implementation of the cascade-crystal configuration is discussed in [16].

## 2. Measurement of qutrits (Bob's station).

The measurement of the arriving qutrits is performed at the Bob's station where he has to measure them in a randomly chosen basis. The problem arising here is that one cannot switch between the proposed states from one to another as they have different polarization degrees. This is so because usage of only linear polarization transformations, like retardation plates does not change the polarization degree. Therefore, one has to use non-linear interactions to switch from one basis to another and this does not seem practical at the moment. The solution is to use the random number generator to decide in which basis we are going to measure our qutrits. Certainly, this is a weak place in the experimental realization of the protocol, because the randomness of the choice is given by the computational but not a physical process. But considering the recent developments in technology, one can build a random number generator where randomness is guaranteed by the physical process and the first devices were already introduced [17]. The second problem that arises here is that even if we have decided about the basis we still have to distinguish the states within these bases from one another with a unity probability. This leads to the problem of filtering the orthogonal polarization states of the biphoton from one another. As it was suggested in [18], it can be done by observing the anticorrelation effect of the general type, i.e., for biphotons with an arbitrary polarization degree in the Brown-Twiss scheme with polarization filters in its arms (Fig. 3). This filter consists of a pair of phase-plates ($\frac{l}{2}$ and $\frac{l}{4}$) and a polarization analyzer. Changing the orientation of the plates relatively to the vertical axis (parameters $q_{1/2}, q_{1/4}$) one can achieve any polarization states of single photons forming the biphoton. Assume that one sets the filters in such a way that in each arm, complete transmittance of the corresponding photon from a pair is observed. In this case, we will be saying that the Brown-Twiss scheme is "tuned" to a certain polarization state of the biphoton. The criterion of the orthogonality of two biphotons is identical to the observation of zero coincidence rate in the case where the input of the Brown-Twiss scheme gets a biphoton that is orthogonal to the one at which the scheme is "tuned". This allows us to distinguish between the orthogonally polarized biphotons that constitute the basis in the protocol from one another making it possible to distill the ternary raw key.

Bob's station can be built as in Fig.3. It consists of three Brown-Twiss schemes with 50% non-polarizing beam-splitters and polarization filters. On the first stage, Bob has to randomly choose the basis in which he will measure the incoming qutrit. In practice, this can be achieved by randomly setting out the certain parameters of the polarization filters in the Brown-Twiss schemes. For example, if we choose to measure in the basis that is given by the states $|a\rangle, |b\rangle$ and $|g\rangle$, then the first scheme is "tuned" to the state $|a\rangle$. It means that filters in each arm transmit horizontal polarization. The second scheme must be set to $|b\rangle$, i.e. the first arm transmits horizontal polarization and the second transmits the vertical one and so on. The incoming qutrit from Alice can appear at the input of each Brown-Twiss scheme with equal probability with the help of the symmetric three-output splitter. By the procedure of basis reconciliation, the unfortunate results, when Bob did not guess the basis correctly, can be eliminated.

The positive side of this setup that makes it more preferable to the other proposed interferometric setups (see [7] for example) is that we are working with only two-armed interferometer that does not require high stability over time and temperature. Moreover, this interferometer is located only in the preparation side of the setup i.e. at Alice's station. We also want to notice that the achievement of higher dimensionality is based on using the specific states of quantum light fields (biphotons as qutrits), rather then increasing the Hilbert space dimension of the single photons. The combination of two methods indicated above may also produce a reasonable and practical resource for the further increase of system's dimension.

Finally we would like to discuss the possible losses which are present in the scheme under consideration. Definitely this scheme suffers from the losses that cannot be in principle eliminated. First of all, Bob may not guess the basis correctly, which leaves us $\frac{1}{4}$ of the number of qutrits. This kind of losses can be considered as inescapable, because it is caused by the nature of the protocol. Then on the stage of the state separation at the three-output beam-splitter one may receive a biphoton at the input of the Brown-Twiss scheme with a probability of $\frac{1}{3}$. Considering the action of the non-polarizing beam-splitter on the biphoton, we are left with $\frac{1}{2}$ of the cases when it may result in a coincidence. And finally, the action of the polarizing filters on the halves of the biphoton may also filter out $\frac{1}{2}$ of the fortunate cases. It is noteworthy to mention that such action of the filters gives us the upper bound of the losses. The lower bound is given when the filters act in the same way on the halves of the biphoton. Then the coincidence rate does not depend on how the biphoton was split at the beam-splitter. Therefore the total losses in a proposed scheme vary from 96% to 98%. Subtracting inescapable losses, arising from the random basis choice (75%) one gets about 8% events of successful qutrit registration by Bob sent in a given basis by Alice. All of them are stipulated by accidental splitting of light on beam-splitters. The presented numbers show that the practical implementation of this protocol will hardly enhance the bit rate of the key transition, but the higher dimensionality of the carriers gives us higher security then the traditional qubit protocols [8, 9].

**Conclusion**

We propose the first concrete experimental setup for the realization of the QC protocol that exploits the polarization properties of single-mode biphotons. The examination of the setup showed that the losses that will be induced by various elements of the setup are quite high for the implementation of such polarization-based systems in practical purpose. The positive side is that it is in principle possible to prepare a polarization-based optical qutrit without any interferometric scheme that requires the stability over time. This can be done in the scheme with three non-linear crystals located in the common pump one after another. This scheme is now being considered. Still the proposed possibility to prepare an optical three-level system, using three non-linear crystals with quite high efficiency and no principal losses at Alice's station seems very attractive for its further investigation.

**Acknowledgments**

This work was supported in part by the Russian Foundation for the Basic Research (02-02-16843, 03-02-16444) and INTAS (grant 2122-01).

**LITERATURE**

1. C.H. Bennett and G. Brassard, in Proceedings of the IEEE international conference on Computers, Systems and Signal Processing, Bangalore, India (IEEE, New York, 1984), p.175.

2. D.Bruss, Phys.Rev.Let. **81**, 3018 (1998).

3. H. Bechmann-Pasquinucci, N. Gisin, Phys.Rev.A **59**, 4238 (1999).

4. N. Gisin, G. Ribordy, W. Tittel and H. Zbinden, Rev. Mod. Phys. **74**, 145 (2002).

5. G. Ribordy, J-D. Gautier, N. Gisin, O. Guinnard, H. Zbinden, Journal of Modern Optics, , **47**, no. 2/3, 517 (2000).

6. C. A. Fuchs, N. Gisin, R.B. Griffiths, C.S. Niu and A. Peres, Phys.Rev.A **56**, 1163 (1997).

7. H. Bechmann-Pasquinucci, W. Tittel, Phys.Rev.A **61**, 062308 (2000 ).

8. H. Bechmann-Pasquinucci, A. Peres, Phys.Rev.Let. **85**, 3313 (2000).

9. D. Bruss, C. Machiavello, Phes.Rev.Lett. **88**, 127901 (2002)

10. D. Kaszlikowski, D.K.L. Oi, M. Christandl, K. Chang, A. Ekert, L.C. Kwek and C.H. Ou, Phys.Rev.A **67**, 012310 (2003).

11. T. Durt, N. J. Cerf, N. Gisin and M. Zukowski, Phys.Rev.A **67**, 012311 (2003).

12. A.V. Burlakov, D.N. Klyshko, JETP Lett. **69**,   11, 795 (1999).

13. A.V.Burlakov, M.V.Chekhova, O.A.Karabutova, D.N.Klyshko, and S.P.Kulik, Phys.Rev.A **60**, No. 6, R4209 (1999).

14. A.V. Burlakov, L.A. Krivitskiy, S.P. Kulik, G. A. Maslennikov and M.V. Chekhova, e-print quant-ph/0207096.

15. A.V. Burlakov, M.V. Chekhova, JETP Lett. **75**, 8, 432 (2002).

16. Y.H.Kim, M.V.Chekhova, S.P.Kulik, M.Rubin, and Y.H.Shih. Phys.Rev.A **63**, 062301 (2001).

17. A. Stefanov, N. Gizin, O. Guinnard, L. Guinnard, H. Zbinden, e-print quant-ph/9907006.

18. A.A Zhukov, G.A. Maslennikov and M.V. Chekhova, JETP Letters,  76, (10),  696 (2002).

**FIGURE CAPTIONS**

Table 1. Amplitudes and phases of the states that are used in the QC protocol.

Fig. 1a Graphical representation of an arbitrary polarization state of the biphoton field on the Poincare sphere ($\vartheta$, $\varphi$ are the azimuthal and polar angles that are used in the standard Poincare sphere technique, $\alpha$ -angle that determines the polarization degree of a biphoton).

Fig. 1b Mapping of the twelve states which are used in the protocol on the Poincare sphere.

Fig.2. Preparation of an arbitrary polarization state of a single-mode biphoton (qutrit) with given parameters. $\left(\lambda/2\right)_{1,2}$ are the half-wave plates to change the amplitudes of the basis states; P.S.1 is the phase shifter between the states $|2,0\rangle$ and $|0,2\rangle$; P.S.2 is the phase shifter of the $|1,1\rangle$ - state with respect to superposition obtained using two type-I crystals; DM is a dichroic mirror that transmits the biphoton field coming from the long arm and reflects the pump coming through the short arm; F is the pump cut-off filters.

Fig.3. Schematic drawing of the Bob's station. RNG is the random number generator that defines the basis in which Bob measures the incoming qutrits; the choice of the state may be done using the symmetric three-output beam-splitter; WP are the wave-plates forming the polarization filters; A are the analyzers; D are the detectors; C.C is the double-coincidence scheme.

| | $|c_1|$ | $|c_2|$ | $|c_3|$ | $f_1$ | $f_2$ | $f_3$ | P |
|---|---|---|---|---|---|---|---|
| $|a\rangle$ | 1 | 0 | 0 | 0 | 0 | 0 | 1 |
| $|b\rangle$ | 0 | 1 | 0 | 0 | 0 | 0 | 0 |
| $|g\rangle$ | 0 | 0 | 1 | 0 | 0 | 0 | 1 |
| $|a'\rangle$ | $1/\sqrt{3}$ | $1/\sqrt{3}$ | $1/\sqrt{3}$ | 0 | 0 | 0 | $\frac{2\sqrt{2}}{3}$ |
| $|b'\rangle$ | $1/\sqrt{3}$ | $1/\sqrt{3}$ | $1/\sqrt{3}$ | 0 | 120 | -120 | $\frac{2\sqrt{2}}{3}$ |
| $|g'\rangle$ | $1/\sqrt{3}$ | $1/\sqrt{3}$ | $1/\sqrt{3}$ | 0 | -120 | 120 | $\frac{2\sqrt{2}}{3}$ |
| $|a''\rangle$ | $1/\sqrt{3}$ | $1/\sqrt{3}$ | $1/\sqrt{3}$ | 120 | 0 | 0 | $\frac{2\sqrt{2}}{6}$ |
| $|b''\rangle$ | $1/\sqrt{3}$ | $1/\sqrt{3}$ | $1/\sqrt{3}$ | 0 | 120 | 0 | $\frac{2\sqrt{2}}{6}$ |
| $|g''\rangle$ | $1/\sqrt{3}$ | $1/\sqrt{3}$ | $1/\sqrt{3}$ | 0 | 0 | 120 | $\frac{2\sqrt{2}}{6}$ |
| $|a'''\rangle$ | $1/\sqrt{3}$ | $1/\sqrt{3}$ | $1/\sqrt{3}$ | -120 | 0 | 0 | $\frac{2\sqrt{2}}{6}$ |
| $|b'''\rangle$ | $1/\sqrt{3}$ | $1/\sqrt{3}$ | $1/\sqrt{3}$ | 0 | -120 | 0 | $\frac{2\sqrt{2}}{6}$ |
| $|g'''\rangle$ | $1/\sqrt{3}$ | $1/\sqrt{3}$ | $1/\sqrt{3}$ | 0 | 0 | -120 | $\frac{2\sqrt{2}}{6}$ |

Table 1.

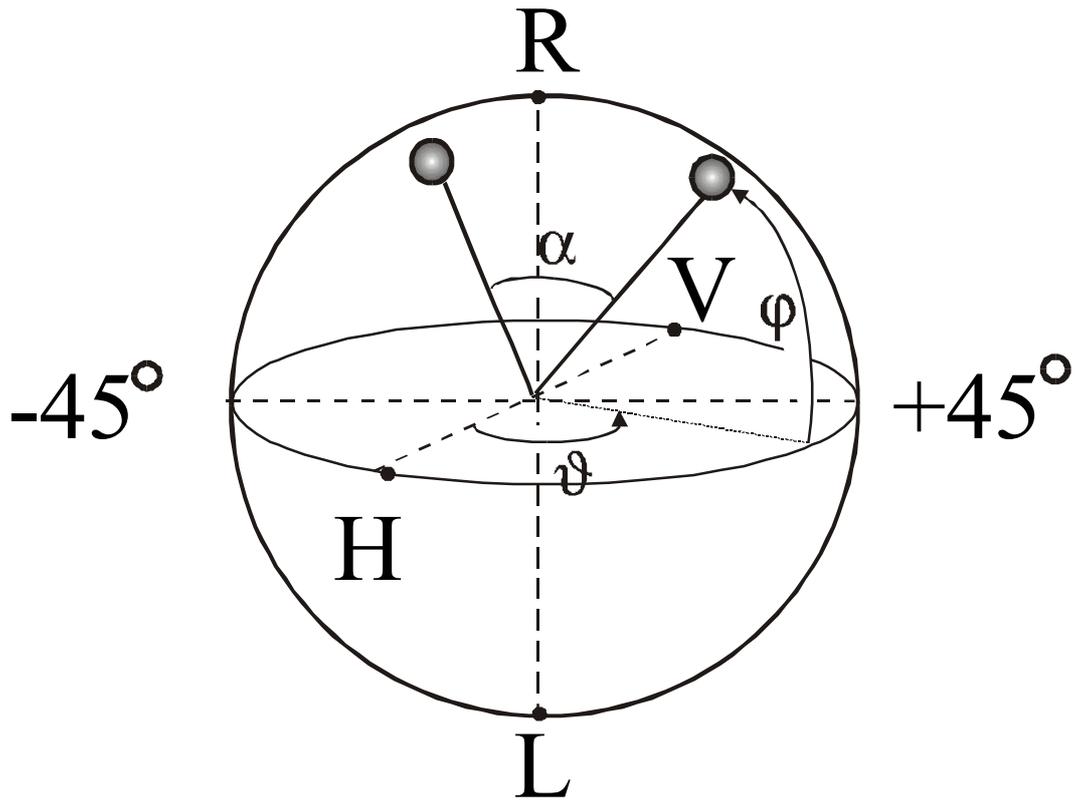

Fig. 1a

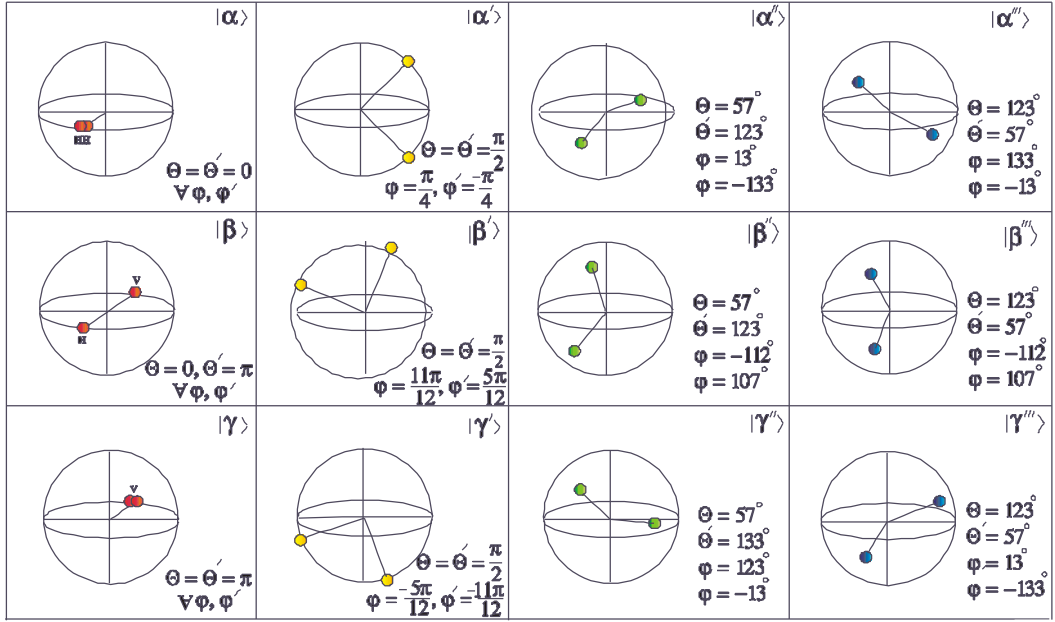

Fig.1b

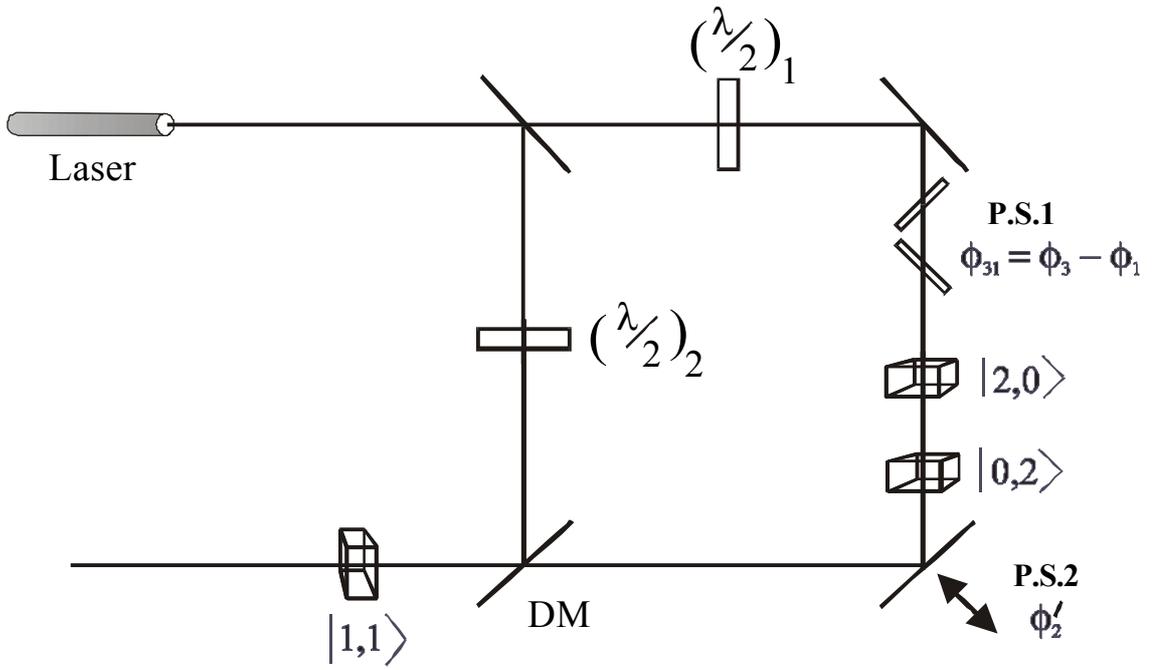

Fig.2

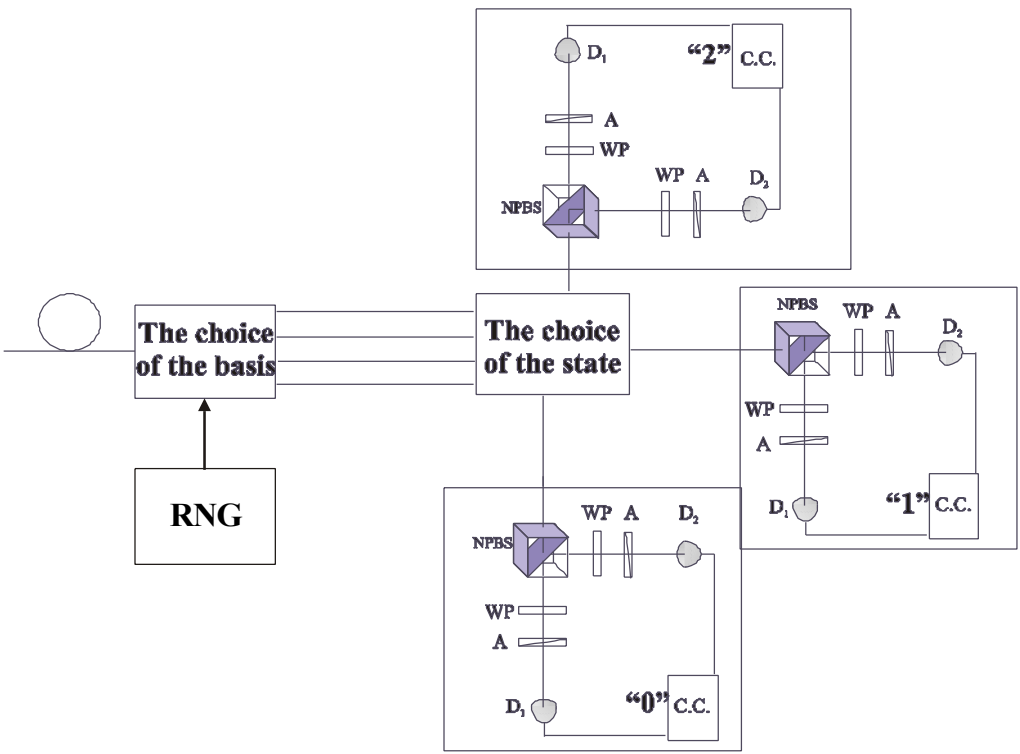

Fig.3